\documentclass[twocolumn]{emulateapj}

\def\kms{${\rm km\,s}^{-1}$}
\newcommand{\degree}{\ensuremath{^{\circ}}}

\shorttitle{Dual Trigger of Prominence Oscillations}
\shortauthors{Gosain and Foullon}

\begin{document}

\title{Dual Trigger of Transverse Oscillations in a Prominence by EUV Fast and Slow Coronal Waves: SDO/AIA and STEREO/EUVI Observations}

\author{S. Gosain}
\affil{National Solar Observatory, 950 N. Cherry Avenue, Tucson 85719, Arizona, USA}

\author{C. Foullon}
\affil{Centre for Fusion, Space and Astrophysics, Department of Physics, University of Warwick, Coventry CV4 7AL, UK}

\begin{abstract}
We analyze flare-associated transverse oscillations in a quiescent solar prominence  on 8-9 September, 2010. Both the flaring active region and the prominence were located near the West limb, with a favorable configuration and viewing angle. The fulldisk extreme ultraviolet (EUV) images of the Sun obtained with high spatial and temporal resolution by the Atmospheric Imaging Assembly (AIA) aboard the {\it Solar Dynamics Observatory}, show flare-associated lateral oscillations of the prominence sheet. The STEREO-A spacecraft, 81.5\degree ahead of the Sun-Earth line, provides on-disk view of the flare-associated coronal disturbances. We derive the temporal profile of the lateral displacement of the prominence sheet by using the image cross-correlation technique. The displacement curve was de-trended and the residual oscillatory pattern was derived. We fit these oscillations with a damped cosine function with a variable period and find that the period is increasing. The initial oscillation period ($P_0$) is  $\sim$28.2 minutes and the damping time ($\tau_D$) $\sim$ 44 minutes. We confirm the presence of fast and slow EUV wave components. Using STEREO-A observations we derive a propagation speed of $\sim$250 \kms for the slow EUV wave by applying time-slice technique to the running difference images. We propose that the prominence oscillations are excited by the fast EUV wave while the increase in oscillation period of the prominence is an apparent effect, related to a phase change due to the slow EUV wave acting as a secondary trigger.
 We discuss implications of the dual trigger effect for coronal prominence seismology and scaling law studies of damping mechanisms.
\end{abstract}

\section{Introduction}
Quiescent solar prominence can be easily seen in chromospheric H-$\alpha$ images as dense and cool sheets of solar plasma protruding above the solar limb. They are typically 100 times denser and 100 times cooler than the surrounding coronal plasma \citep{Tandberg1974}. The chromospheric off-limb images reveal their complex structure as thin vertical threads often with oscillatory motions \citep{Ballester2006,Okamoto2007}. One of the earliest studies of the oscillatory phenomena in prominence was done during 1930s \citep{Newton1935}. Since then the field has grown and the prominence oscillations, as observed, are categorized into small amplitude and large amplitude oscillations. The former corresponds to amplitudes of about 2-3 \kms while the latter correspond to
amplitudes in excess of 20 \kms. Generally the large amplitude oscillations are attributed to the powerful  flares and associated Coronal Mass Ejection (CME)
phenomena in the neighborhood of the prominence \citep{Ramsey1966,Eto2002}. However, small-scale jets or
sub-flares in the neighborhood have also been found to excite the large amplitude oscillations \citep{Isobe2007,Vrsnak2007}. In some cases the oscillations are also accompanied
during filament eruptions \citep{Isobe2006,Gosain2009, Artzner2010}.

The oscillatory phenomena in prominence can be observed in two
types of measurement or via two types of technique.
(i) {\it Spectroscopic mode:}  which gives line-of-sight (LOS) component of the velocity of motion. Such observations are generally obtained with tunable filters or slit spectrographs. The filament appears and disappears
periodically in the images obtained with the filter tuned to the H-$\alpha$ line center. This is termed as the 'winking' of the filament.
(ii) {\it Imaging mode:} where the proper motion of the prominence can be tracked in the plane of sky (POS). This technique gives the horizontal component of the velocity of motion. Using a combination of both techniques \cite{Schmieder2010} could derive the velocity vector in the prominence observed on the limb. Similar techniques were used by \cite{Isobe2006} using H-$\alpha$ dopplergrams and EUV images.

During the last two decades, high-resolution observations of the solar coronal dynamics became available with the advent of space missions, namely, the {\it Solar and Heliospheric Observatory} (SOHO), the {\it Transition Region and Coronal Explorer} (TRACE),
and more recently {\it Hinode} and the {\it Solar Dynamics Observatory} (SDO). The detection of large amplitude and long-period oscillations in prominence, in particular, has been revealed in greater detail by space-based EUV observations from the Extreme ultraviolet Imaging Telescope (EIT)
on board SOHO \citep{Foullon2004,Foullon2009,Hershaw2011} and more recently the Atmospheric Imaging Assembly (AIA) on SDO \citep{Asai2012,Liu2012}.

The flaring regions sometimes produce large-scale disturbances, such as Moreton waves seen in chromospheric H-$\alpha$ filtergrams \citep{Moreton1960} or coronal disturbances seen in EUV images, also known as EUV waves \citep{Thompson1999}. It was earlier thought that the Moreton waves are nothing but chromospheric footprints of dome-shaped magnetohydrodynamic (MHD) wavefronts \citep{Uchida1968}, detected as coronal EUV waves \citep{Thompson1999}. However, later studies showed a number of differences \citep{Klassen2000,Warmuth2007,Thompson2009}. More recent EUV observations with higher spatial and temporal resolution from SDO/AIA suggest a two component structure of coronal EUV disturbances, i.e. a faint fast EUV wave ahead of the slower bright EUV wave \citep{Chen2011,Asai2012,Liu2012}.


The coronal EUV disturbances can interact with distant filament/prominences and excite large amplitude oscillations in them.
In the present study we analyze transverse large scale prominence oscillations associated to a nearby flaring active region and investigate the effect of the two component EUV coronal wave structure. The paper is organized as follows.  Section 2 describes the observations. In Section 3, we present various methods of analysis and their results, and finally we discuss these results in Section 4.

\section{Observations}

\begin{figure}    
\centerline{\includegraphics[width=1.0\textwidth,clip=]{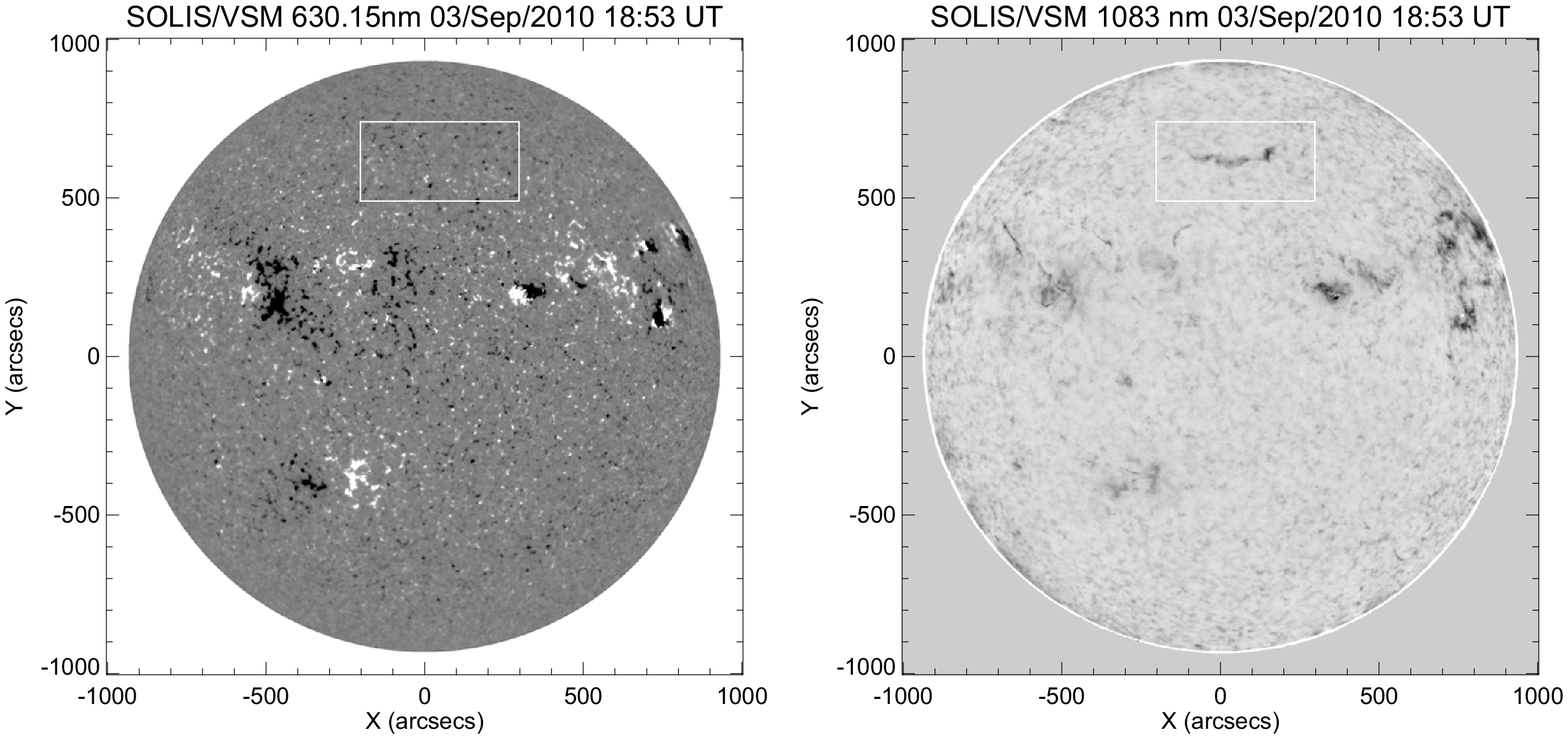}}
\centerline{\includegraphics[width=0.65\textwidth,clip=]{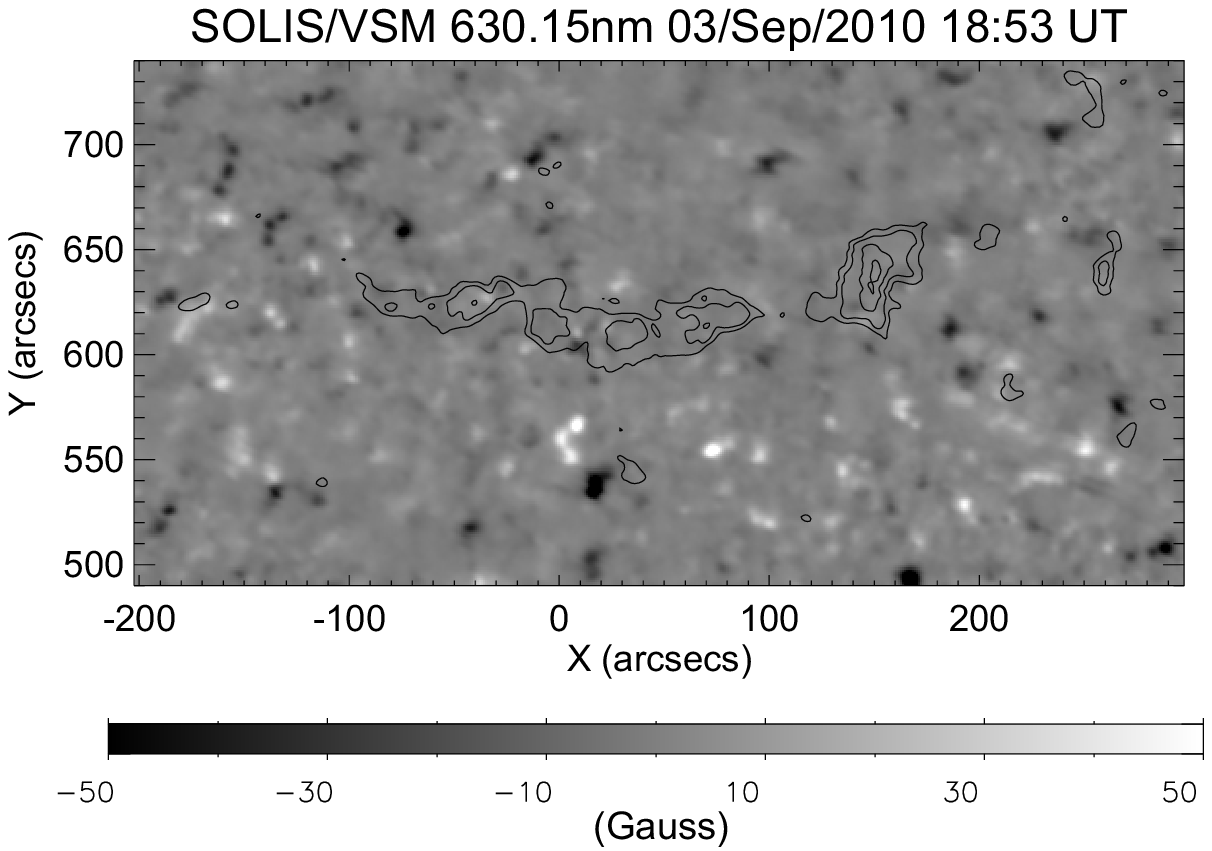}}
\caption{The top panels show context fulldisk observations of the photospheric longitudinal magnetic flux (left panel) and the (He \textsc{I}) 1083 nm image (right panel)  during 3-Sep-2010 from VSM/SOLIS instrument.  The magnified view of the region-of-interest (the filament) is shown in the bottom panel. The contours of 1083 nm absorption are overlaid on the longitudinal magnetogram.  The colorbar at the bottom shows the scaling of the magnetogram pixels between $\pm$ 50 G for both the fulldisk (top left panel) and the limited field-of-view (bottom panel) magnetogram.}
\label{solis-context}
\end{figure}

On 8-9 September, 2010, the prominence and flaring active region  NOAA 11105 were both located on the west solar limb (as observed from the Earth) with a favorable configuration and viewing angle for observing the lateral oscillations of the prominence sheet. The flare detected by the {\it Geostationary Operational Environmental Satellite}-14 (GOES-14) \citep{Hanser1996} is a C3.3-class flare starting at 23:05 UT and peaking at 23:33 UT on 8 September, as measured in the 1-8 \AA~ channel. Figure \ref{solis-context} shows the context full-disk magnetogram and (He \textsc{I}) 10830 \AA\ images taken  on 3 September, 2010, by the Vector Spectromagnetograph (VSM) of the {\it Synoptic Optical Long-term Investigations of the Sun} (SOLIS) facility \citep{Keller2003}.
 The quiescent filament was then crossing the central meridian in the Northern solar hemisphere (inside the white rectangular box, top panel) and appeared as a diffuse elongated structure with weak magnetic polarities ($<$~50 G) in the filament channel (lower panel of the Figure~\ref{solis-context}).

\begin{figure}    
\centerline{\includegraphics[width=0.75\textwidth,clip=]{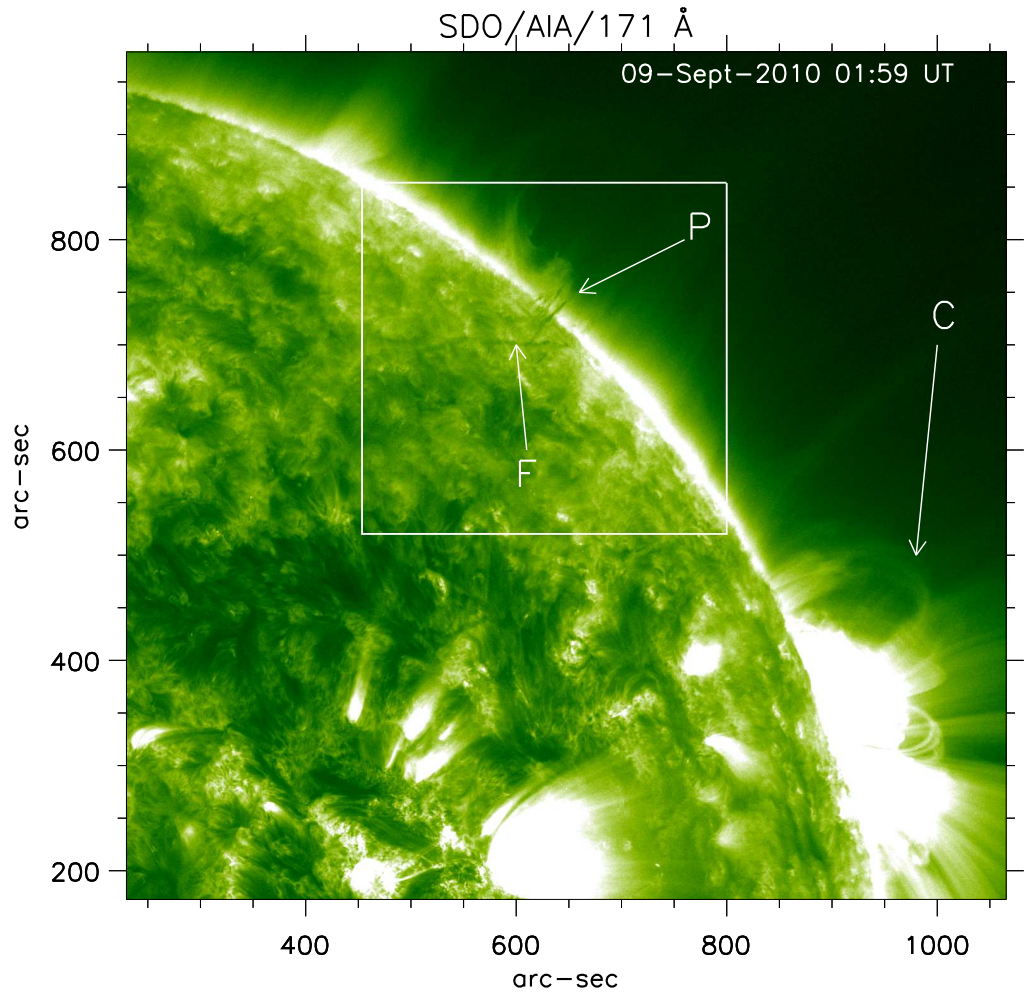}}
\caption{Portion of the solar disk imaged by SDO/AIA instrument in Fe IX 171 \AA.
The white rectangle indicates the FOV used for more detailed image analysis.
The white arrow  labelled `P' points to the part of the filament seen above the limb as a prominence. The arrow labelled `F' locates the dark spine of the filament seen against the solar disk in absorption. The arrow labeled `C' points towards the loops in the active region NOAA 11105, where the flare occurred.}
\label{aia-context}
\end{figure}

We use high-resolution filtergrams in the (Fe \textsc{IX}) 171 \AA~ wavelength observed by the SDO/AIA instrument \citep{Lemen2012}.
SDO/AIA  provides high spatial resolution observations over its large field-of-view (FOV) of 1.3 solar diameters. The spatial sampling is 0.6 arcsec per pixel and the cadence is about 12 seconds. The dataset used in the present study consists of cutout images with a FOV of 836 $\times$ 806 arcsec square centered at heliocentric coordinates (648",576") in the North-West quadrant.  Figure \ref{aia-context} shows one such AIA (Fe \textsc{IX}) 171 \AA\ image, for context, on 9 September 2010, when the filament was reaching the west limb. We can see the filament spine (`F');
the part of the prominence is seen above the limb against the dark sky (`P');
the active region where the  flare and linked CME occurred is also seen at the limb (`C').
The movie of the flare induced filament oscillations is made available as an online material (movie1.mpg).
The white rectangle in Figure~\ref{aia-context} marks the region-of-interest (ROI) within the FOV that we use for further analysis.
This ROI is rotated so as to align the solar limb vertically and to detect maximal displacement along  this preferred direction, as shown in Figure~\ref{time-slice} (top panel, in inverted levels, with emission increasing from bright to dark). The apparent height $H$ of the prominence above the solar limb is $\sim$36 Mm.

To deduce the properties of the flare-associated EUV disturbances that triggered the prominence oscillations, we also use (Fe \textsc{XII}) 195 \AA~ observations from the
 Extreme Ultraviolet Imager \citep[EUVI;][]{Wuelser2004} aboard the {\it Solar Terrestrial Relations Observatory Ahead} \citep[STEREO-A;][]{Kaiser2008} (part of the {\it Sun Earth Connection Coronal and Heliospheric Investigation} \citep[SECCHI;][]{Howard2008}) as well as radio observations from the STEREO/WAVES instrument \citep{Bougeret2008}. Other SDO/AIA observations of the EUV disturbances for the same event have been presented by \cite{Liu2012}.
In Figure~\ref{stereo-context} we show the  EUVI observations of the filament in (He \textsc{II}) 304 \AA\ as observed from STEREO-A. STEREO-A was $\sim$81.5$^\circ$ ahead of the Earth, which enabled it to observe the on-disk view of the prominence as well as the flaring region (indicated by arrow). Figure~\ref{stereo-context} shows the image in grayscale for better contrast.  The length of the filament (2L) is estimated to be $\sim400$ Mm.

\section{Analysis of the Oscillations}

\subsection{Space-Time Plot of Prominence Threads}

\begin{figure}    
\centerline{\includegraphics[width=0.8\textwidth,clip=]{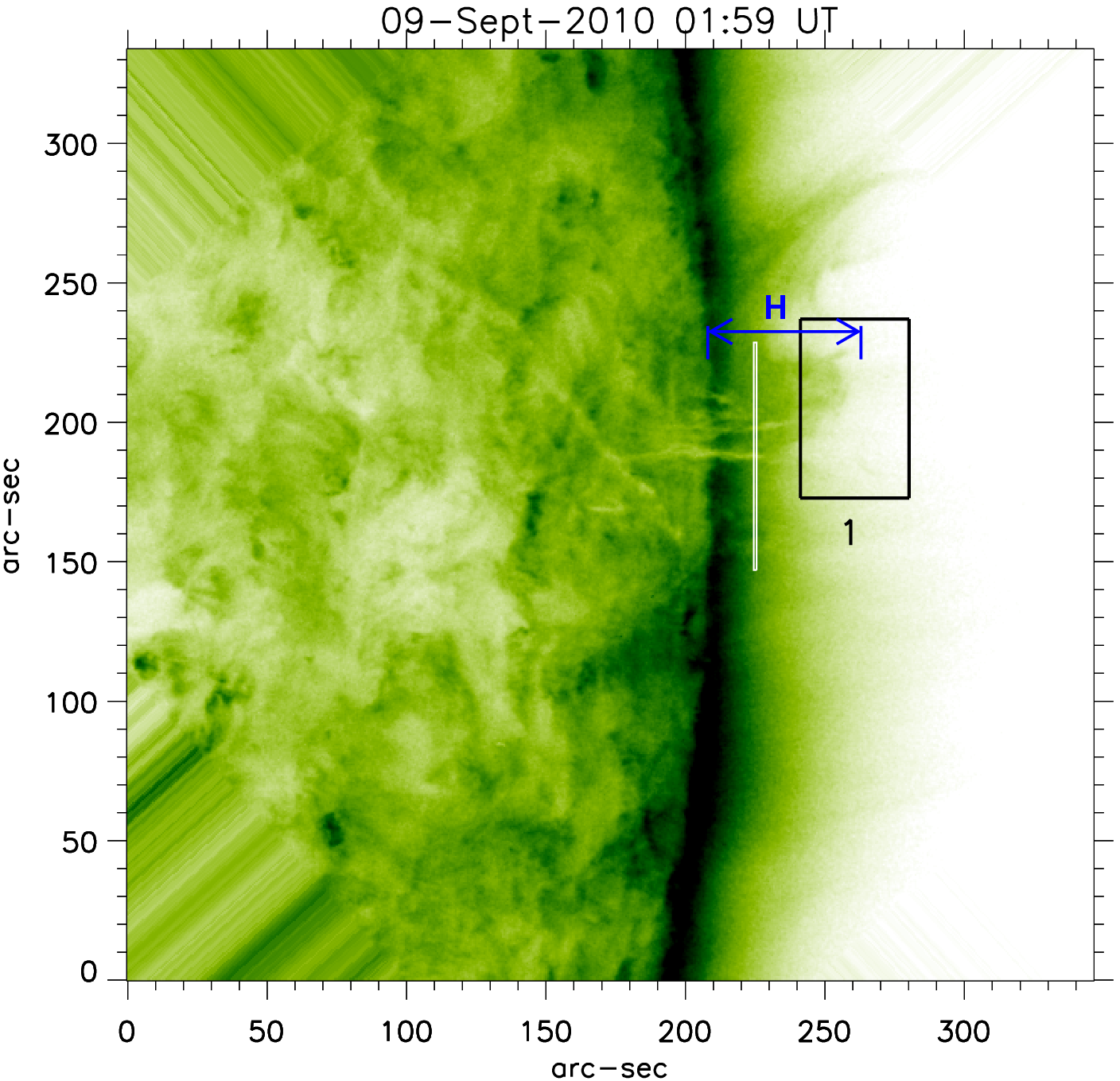}}
\centerline{\includegraphics[width=0.8\textwidth,clip=]{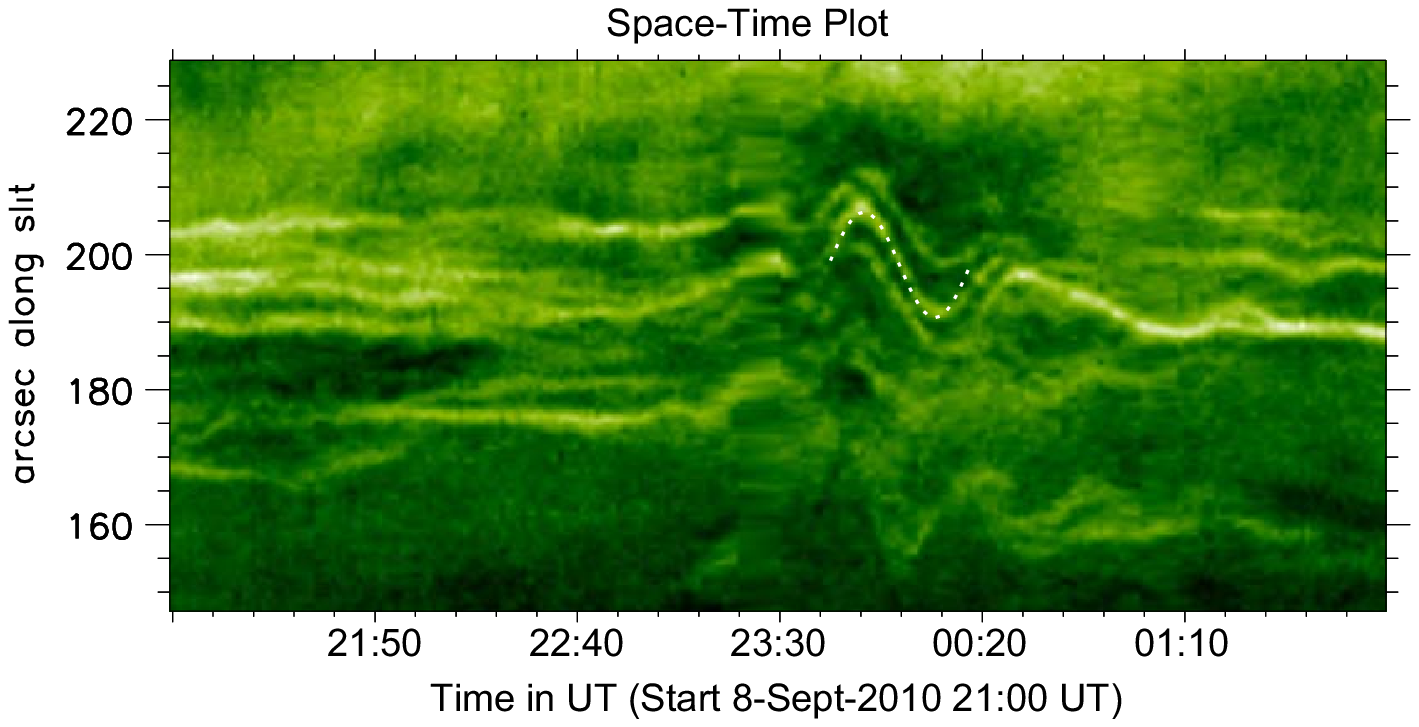}}
\caption{(Upper) Image obtained after rotation of the FOV, indicated by a white rectangle in Figure \ref{aia-context}, aligning the filament threads along the horizontal axis; the black rectangular box labeled `1' is used for tracking the oscillations in the top part of the prominence by performing cross-correlation analysis of subsequent frames; the height $H$ of the  prominence above the solar limb is $\sim~$36 Mm.
(Lower) Space-time diagram constructed from the artificial slit indicated by the white line in the upper image; a white dotted sinusoid corresponds to a period of about 28 minutes.}
\label{time-slice}
\end{figure}

\begin{figure}    
\centerline{\includegraphics[width=0.8\textwidth,clip=]{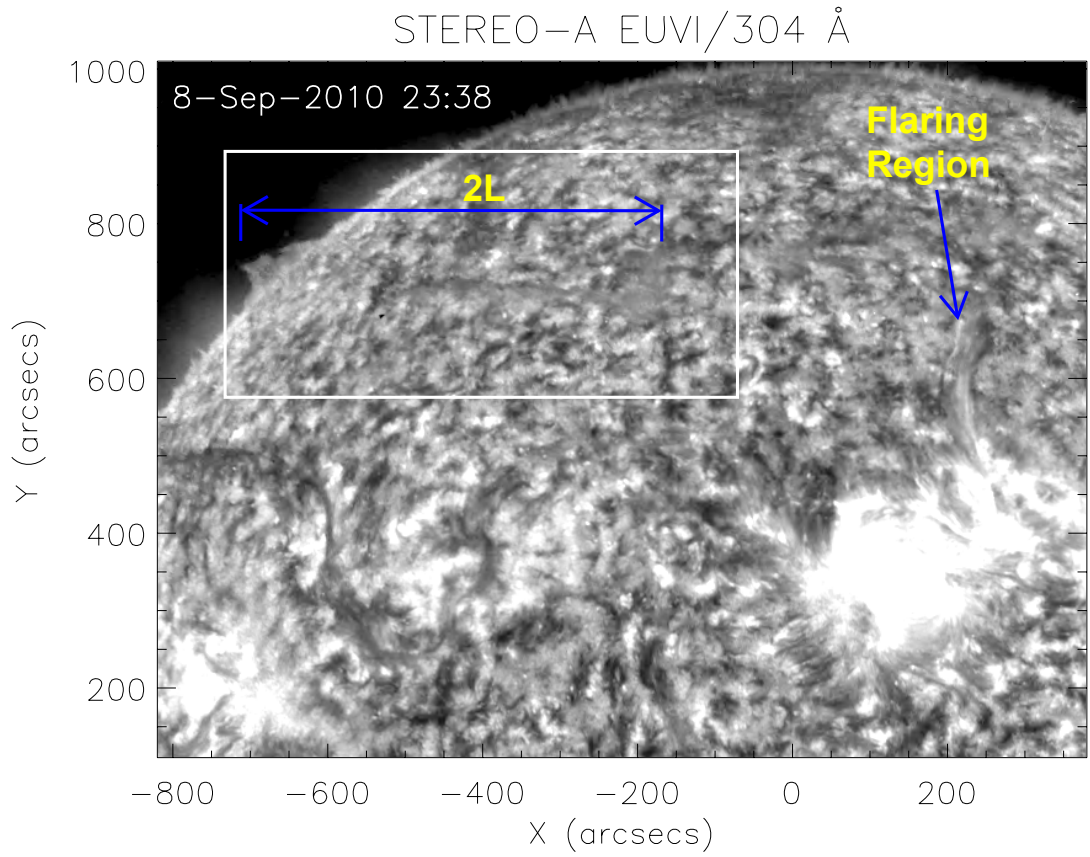}}
\caption{ Image in the (He \textsc{II}) 304 $\AA~$ wavelength by STEREO-A/EUVI, showing in the Northern hemisphere the flaring region and (inside a white rectangular box) the filament. The image is displayed in gray scale for higher contrast. The filament length ($2L$) is $\sim$400 Mm.}
\label{stereo-context}
\end{figure}

 The observations (movie1.mpg) suggest that the oscillations of the prominence sheet are predominantly lateral. The white line segment drawn in the top panel of Figure~\ref{time-slice} represents the artificial slit along which we track the time evolution of prominence threads. To analyze the oscillatory motion of the filament threads in
response to the flare-associated disturbances, we make space-time plot as shown in the bottom panel of Figure~\ref{time-slice}.
It can be seen from the time-slice image (also from the movie, movie1.mpg) that the threads evolve as the prominence oscillates.  Initially, the prominence appears to oscillate coherently, as a collection of threads, but progressively threads become out of phase with each other, and begin to loose their identity as they evolve and overlap each other during the oscillations.
The collective transverse oscillation, to a first approximation, is indicative of a global kink mode and the non-collective behaviour can be attributed to the filamentary and non-homogeneous
structure of the prominence \citep{Hershaw2011}.
It may be also noted that before the flare onset, the threads were showing small amplitude oscillatory motions. The periods of these small amplitude oscillations are difficult to measure in the presence of evolving and overlapping threads. Similar small amplitude oscillations were found by \cite{Okamoto2007} using high resolution Ca \textsc{II} H filtergrams obtained by the Solar Optical Telescope \citep[SOT;][]{Tsuneta2008} on board {\it Hinode} \citep{Kosugi2007}.

Here, our focus  is on the flare induced large amplitude oscillations. In the lower panel of Figure~\ref{time-slice} we overplot a sinusoid (white-dotted curve) over the most visible oscillating thread and infer a period $P\sim$28 minutes and an amplitude of $\sim$3 Mm.  In comparison with similar flare-induced oscillatory prominence events, the period $P$ (28 minutes) is intermediate in ranking order between the 15-min period reported by \cite{Asai2012} and the (so far maximum) 99-min period reported by \cite{Hershaw2011}.

\subsection{Local Correlation Tracking Velocity Map}

\begin{figure}    
\centerline{\includegraphics[width=0.8\textwidth,clip=]{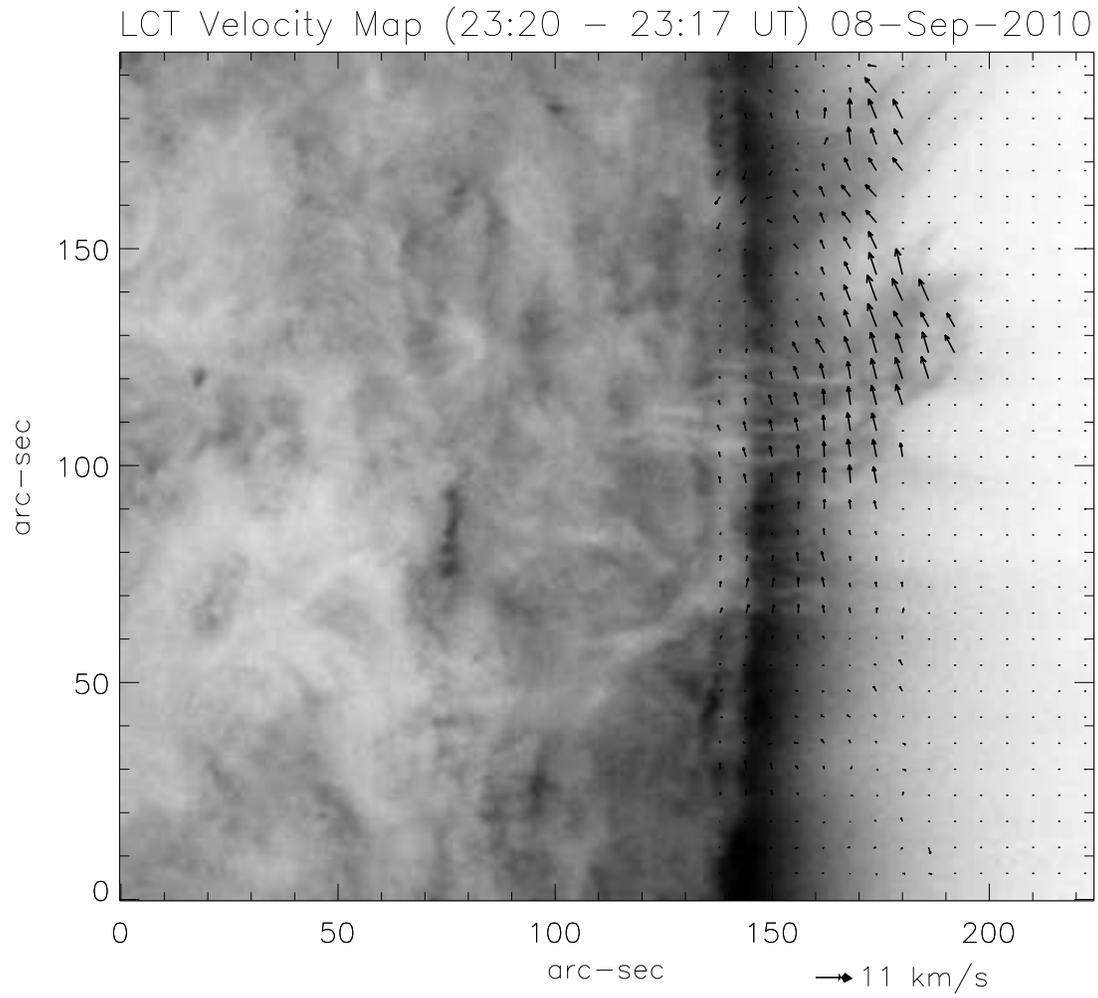}}
\caption{Map of POS velocity vectors derived by applying LCT to the SDO/AIA frames observed between 23:17 and 23:20 UT on 8 September, 2010. Arrowheads point in the direction of motion of the prominence and the length of the arrow corresponds to the velocity  magnitude.
The maximum amplitude of the velocity is $\sim$11 \kms near the top part of the prominence.}
\label{lct}
\end{figure}

We apply the local correlation tracking (LCT) technique \citep{November1988} to the  AIA images taken between 23:17 and 23:20 UT on 8 September, 2010. During this time interval, the prominence sheet undergoes its first oscillation cycle towards its maximal displacement.  We apply the LCT method to the portion of the prominence sheet visible above the solar limb. The LCT method is commonly used on-disk with photospheric observations, such as visible light continuum images or magnetograms. In the case of coronal observations, LCT is more adequately used for deriving the motion vectors in off-limb coronal structures visible against stable sky background.
The LCT map is shown in Figure~\ref{lct}.
The direction of the arrows represent the direction of motion of the prominence sheet locally. The
length of the arrow represents the velocity amplitude. It may
be noticed that: (i) the prominence as a whole moves laterally, as indicated by the uniform direction pointed by most of the arrows, and  (ii) the top of the prominence shows maximum velocity amplitude of about 11 \kms and the lower parts showing relatively smaller velocity amplitude of about 5 \kms. 
 The dependence of velocity amplitude on height, also found previously by \citet{Hershaw2011}, is a characteristic signature of a fundamental oscillatory mode. These results are consistent with the interpretation that the main oscillation is a global kink mode oscillation.
The LCT vectors also show that apart from the lateral (parallel to the solar limb) displacement, there is a downward component as well.

\subsection{Image Cross-correlation Analysis}

\begin{figure}    
\centerline{\includegraphics[width=0.9\textwidth,clip=]{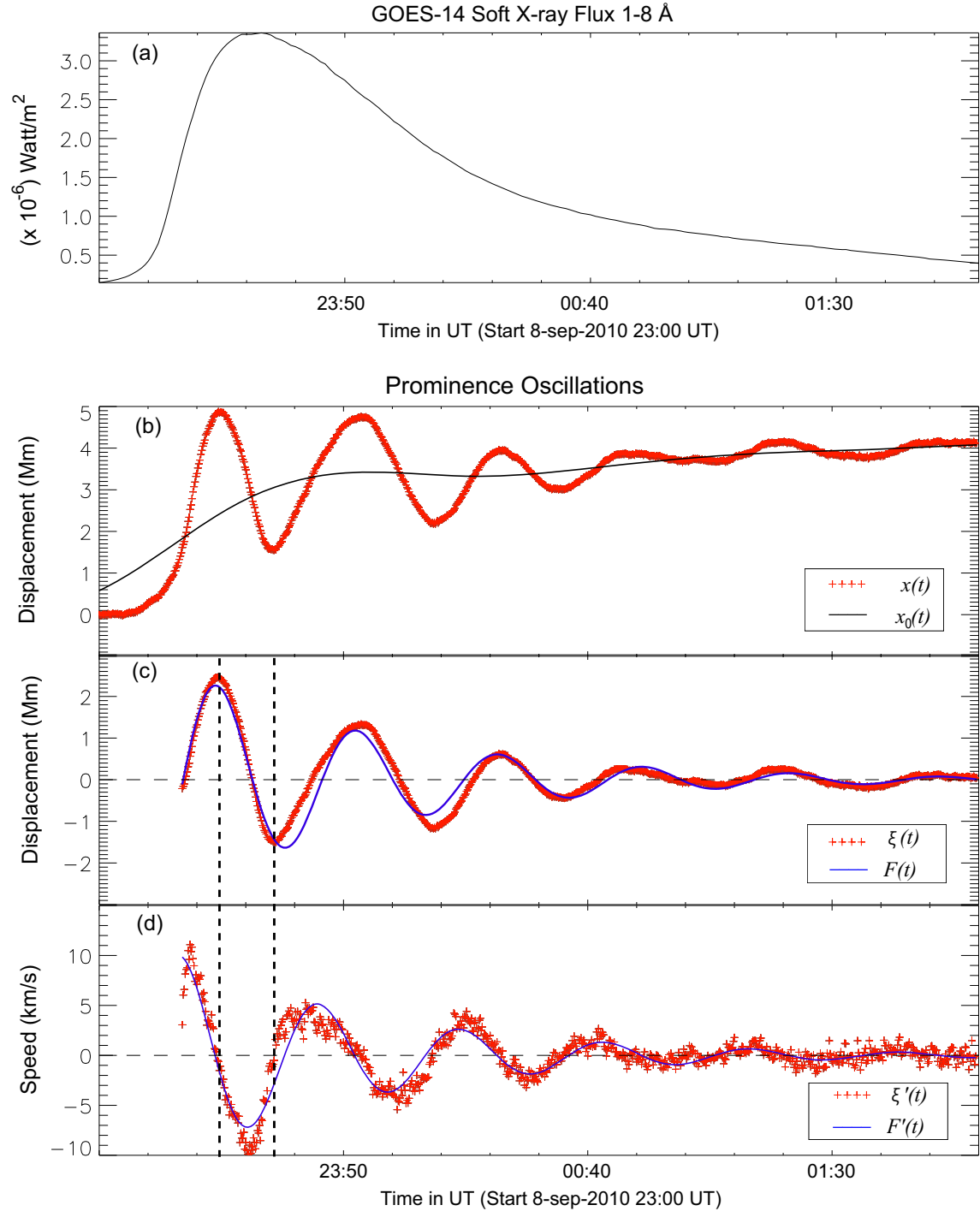}}
\caption{(a) Soft X-ray flux in the 1-8 \AA~ channel of the GOES-14 satellite.
(b) Displacement curve, $x(t)$, (red + symbols) obtained in the top part of the  prominence (box `1' in Figure \ref{time-slice}) and fitted long-term trend of the prominence displacement,  $x_{0}(t)$,  (solid black line).
(c) De-trended displacement profile, $\xi(t)=x(t)-x_{0}(t)$, (red `+' symbols) and fit to a damped cosine curve, $F(t)$ (solid blue line).
(d) Time derivatives of the $\xi(t)$ (red `+' symbols) and $F(t)$ (solid blue line). In panels (c-d), the two vertical dashed lines mark the times 23:25 UT and 23:35 UT, corresponding to the effects of primary and secondary triggers of the oscillations,  respectively. }
\label{displ-curve}
\end{figure}

The LCT analysis done above shows that the maximal displacement occurs near the  top-height part of the prominence. We thus choose the corresponding ROI, marked by a black rectangular box (labelled as `1') in the top panel of Figure~\ref{time-slice}, to perform a  cross-correlation analysis. The following steps are involved: (i) the time sequence of the ROI is first extracted, (ii) the subsequent frames are cross-correlated with respect to the preceding frame and the relative shifts, which maximize cross-correlation coefficient, are derived, and finally (iii) these instantaneous shifts are added to derive the displacement of the prominence, $x(t)$, along the direction normal to the prominence sheet, as shown in Figure~\ref{displ-curve}(b) (red '+' symbols). In Figure~\ref{displ-curve}(a) we show the light curve of the GOES-14 soft X-ray flux in the channel 1-8 \AA\ during the same time interval.

The displacement curve, $x(t)$, exhibits an oscillatory pattern over a non-oscillatory net displacement of the prominence. The trend in the prominence position suggests that the prominence shifts its mean position (by $\sim 4$ Mm in 3 hours) as a whole while exhibiting oscillatory motion.  However, part of this long-term change may be attributed to the effect of solar rotation. More precisely, it can be shown that, with the solar heliographic latitude $\sim 7.2\degree$, about a quarter ($\sim 1.1$ Mm) of the $\sim 4$ Mm apparent latitudinal motion corresponds to solar rotation while the rest of the motion could be a result of a slight latitudinal tilt in the filament orientation itself or a proper motion of the filament. Nevertheless, the mean trend in the displacement of the filament due to various contributing factors needs to be removed before the oscillatory signal can be analyzed further. Following \cite{Foullon2010} we first use the wavelet $\grave{a}$ trous transform \citep{Starck2002} to estimate the mean trend, $x_{0}(t)$, in the displacement profile $x(t)$. This trend shown as a black curve in Figure~\ref{displ-curve}(b), can be understood as the sum of a (secular) quasi-linear component, which can be associated with a solar rotation effect, and an oscillatory component associated with the prominence oscillation. Therefore, we understand to have corrected those effects in the de-trended  oscillatory displacement curve given by $\xi(t)=x(t)-x_{0}(t)$, as shown in Figure~\ref{displ-curve}(c).

From the visual inspection of the de-trended displacement curve, $\xi(t)$, we notice that the oscillation period for the first and second oscillation is not identical, where the second oscillation seems to have slightly larger period than the first oscillation.  We therefore attempt to fit a damped cosine function with a variable period, $F(t)=A cos(2\pi t/(P_0+\lambda t) -\phi) exp(-t/\tau_d)$ to the de-trended displacement curve, $\xi(t)$, and derive amplitude $A$, initial period $P_0$, linear change rate of period $\lambda$, phase $\phi$ and damping time $\tau_d$.
The fitted cosine function is shown by the blue curve in Figure~\ref{displ-curve}(c). The fitted parameters are $A=2.654\pm 0.004$  Mm, $P_0=28.230\pm0.007$ minutes, $\tau_d=43.9\pm0.1$ minutes, and $\lambda=0.0085\pm0.0001$. The small positive value of $\lambda$ suggests that the period is increasing slightly here. For a fitting function without period change ($\lambda$ taken to be zero), one obtains a larger period of $28.94\pm0.07$ minutes. However, despite achieving a better fit to the data with a variable period, the displacement curve departs noticeably from the fitted (blue) curve.

Figure~\ref{displ-curve}(d) shows the time derivative of the displacement curve and its smooth fit, where we have smoothed the curves before taking derivative to reduce the noise amplification. A three point boxcar running average of the displacement curve was taken for smoothing.  This gives us the instantaneous speed curve. The maximum amplitude of the lateral oscillation speed is $\sim$ 11 \kms near 23:18 UT. This is in agreement with the maximum amplitude of the speed derived from LCT analysis in images between 23:17 and 23:20 UT.

Two vertical dashed lines  are drawn in  Figure~\ref{displ-curve}(c-d) to indicate times, first at 23:25 UT, when the displacement amplitude of the first oscillation cycle reaches its maxima, and, second at 23:35 UT, when the displacement curve begin to depart noticeably from the fitted (blue) curve. In the latter case, the departure from the model fit to the data can be explained as a phase change of the ongoing oscillation.

\subsection{EUV Wave Analysis and Dual Trigger Mechanism}

\begin{figure}    
\centerline{\includegraphics[width=1.\textwidth,clip=]{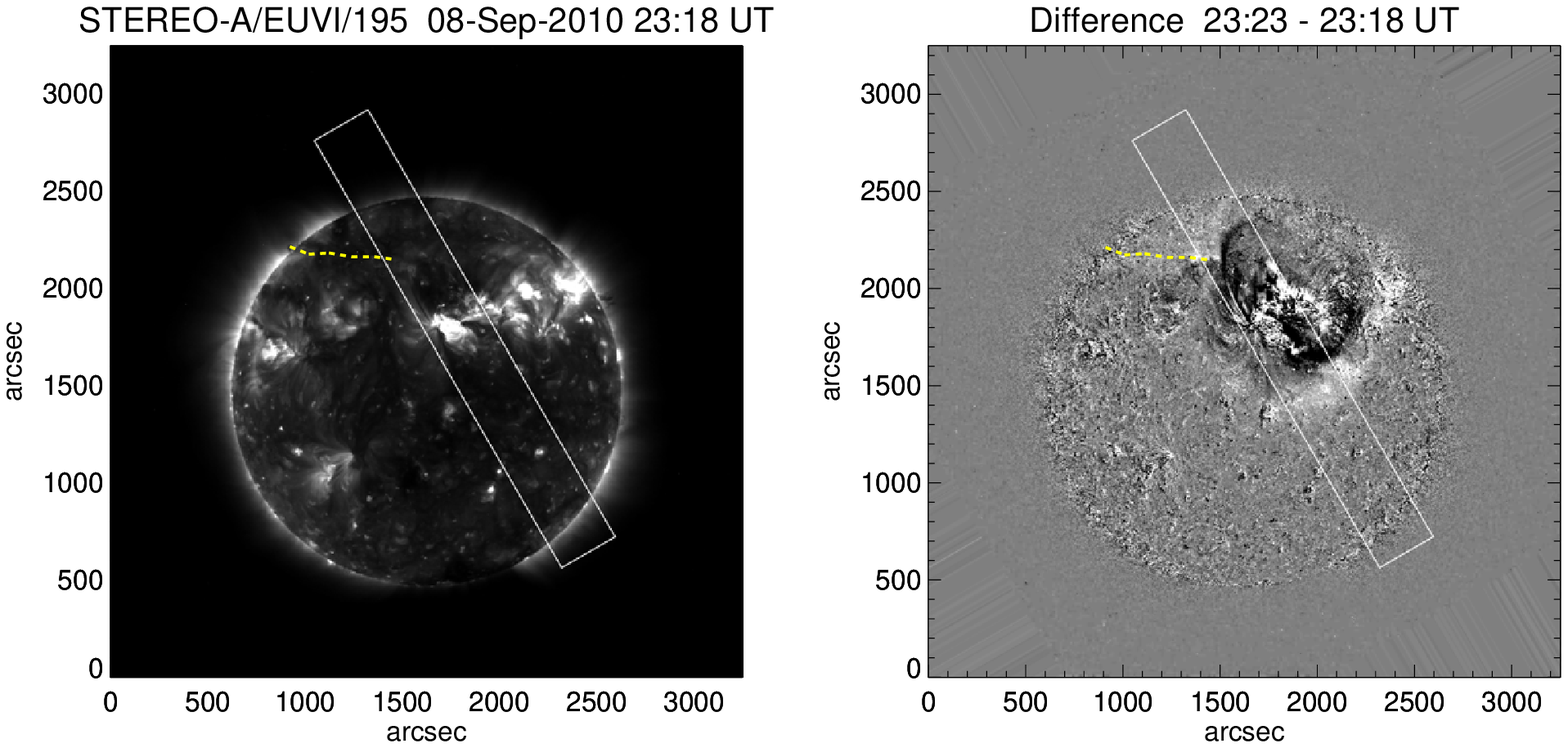}}
\centerline{\includegraphics[width=0.65\textwidth,clip=]{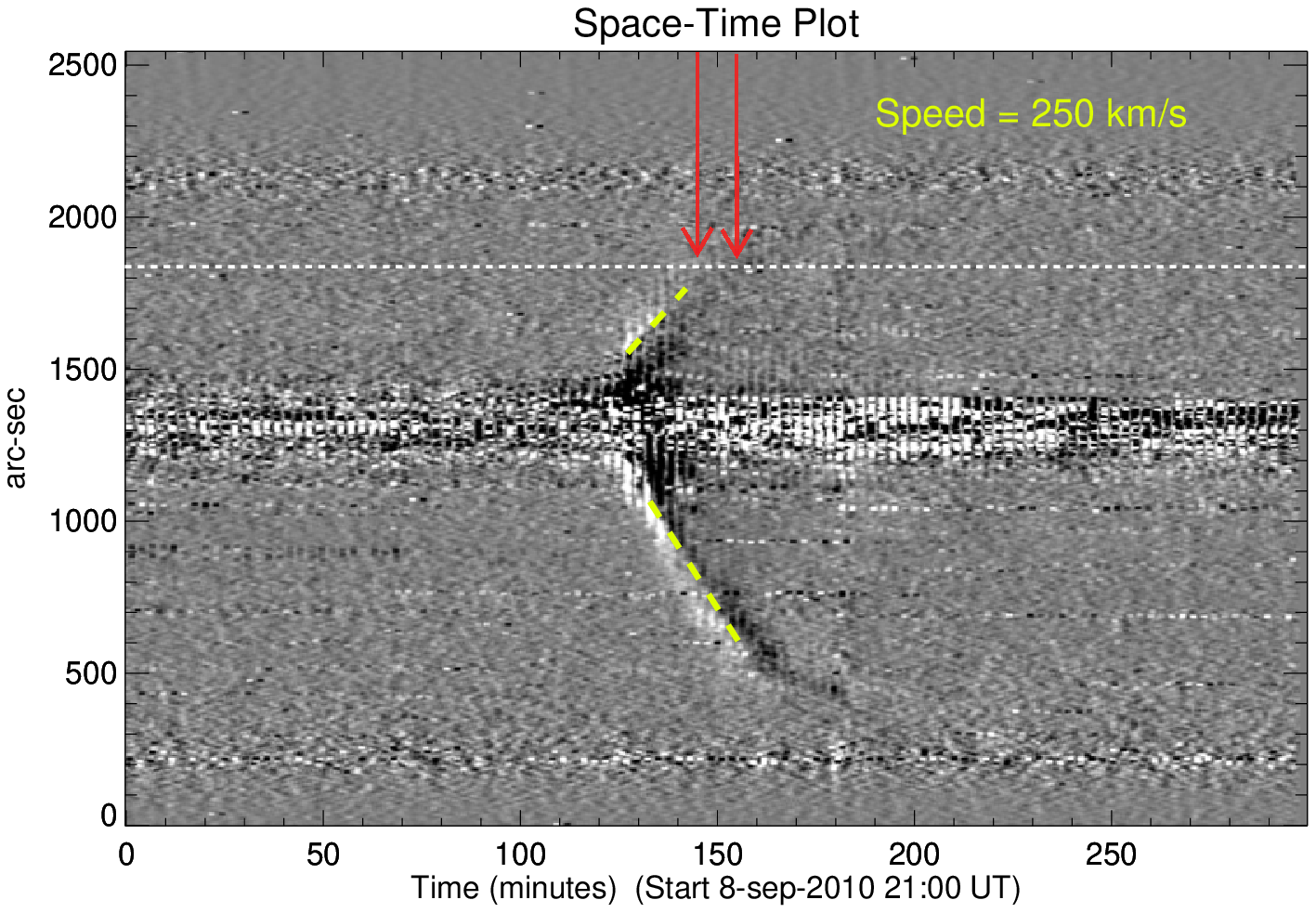}}
\caption{(Upper) STEREO-A 195 \AA~\ images shown as (left) fulldisk and (right) running difference; the white dashed line delineates the position of the filament. (Lower) Space-time diagram constructed from stacking averages of the tilted rectangle box (marked in upper panels) along the narrow side; the yellow dashed line segment marks the slope of EUV wave and gives a POS velocity estimate of about 250 \kms; the horizontal white dashed line indicates the position of the filament; red arrows mark the effects of the primary and secondary triggers of the prominence oscillation at 23:25 UT and 23:35 UT.}
\label{euv-wave}
\end{figure}

The STEREO-A satellite observed the event on-disk due to its large separation angle, i.e., 81.5$^\circ$ ahead of the Earth.  The top left panel of  Figure~\ref{euv-wave} shows the
full-disk EUVI image in 195 \AA. The corresponding running difference image is shown in the top right panel of Figure~\ref{euv-wave}. The expanding EUV wave could be seen in the running difference image quite clearly. The filament, which oscillates on interaction with the flare-related disturbances, is  delineated with a white dashed line in the North-East part of the full-disk images. The rectangular box marked in the top panel images represents the artificial slit used to construct the space-time plot in order to study the EUV wave propagation. The box is positioned such that it adequately covers the flaring active region and is oriented towards the filament.  The space-time plot is shown in the bottom panel of Figure~\ref{euv-wave}. The yellow dashed line marks  the  high-contrast difference signal (i.e., the center of black-white intensity difference pattern)  created by the EUV wavefront propagating Northwards towards the filament. Its slope  corresponds to a velocity of $\sim$250 \kms.   This velocity is projected in the POS and is in the range of the observed values of 200-400 \kms \citep{Klassen2000,Thompson2009,Veronig2010}, typical of a slow EUV wave.  The two red arrows correspond to the times marked in Figure~\ref{displ-curve}(c), indicating the time of first maxima of $\xi(t)$ and the time of change of phase.

An analysis of the EUV wave observed on the limb by SDO/AIA for this event \citep{Liu2012} shows evidence for fast and slow EUV wave components, as well as quasi-periodic wavetrains (with a dominant 2-min periodicity) within the broad EUV pulse. The fast EUV wave component is reported to move at an initial velocity of 1400 \kms decelerating to 650 \kms at a distance greater than $0.5\rm{R}_\odot$, while the velocity of the slow component is found to be around 240$\pm$20 \kms, in agreement with the value derived from our STEREO-A observations.
The presence of fast and slow wave can be visually distinguished in the AIA running difference images \citep[][ their figure 5(l)]{Liu2012}. The separation of the fast and slow wavefronts is about 200 arcsec, i.e. 150 Mm in the direction of the prominence.

\begin{figure}    
\centerline{\includegraphics[width=0.95\textwidth,clip=]{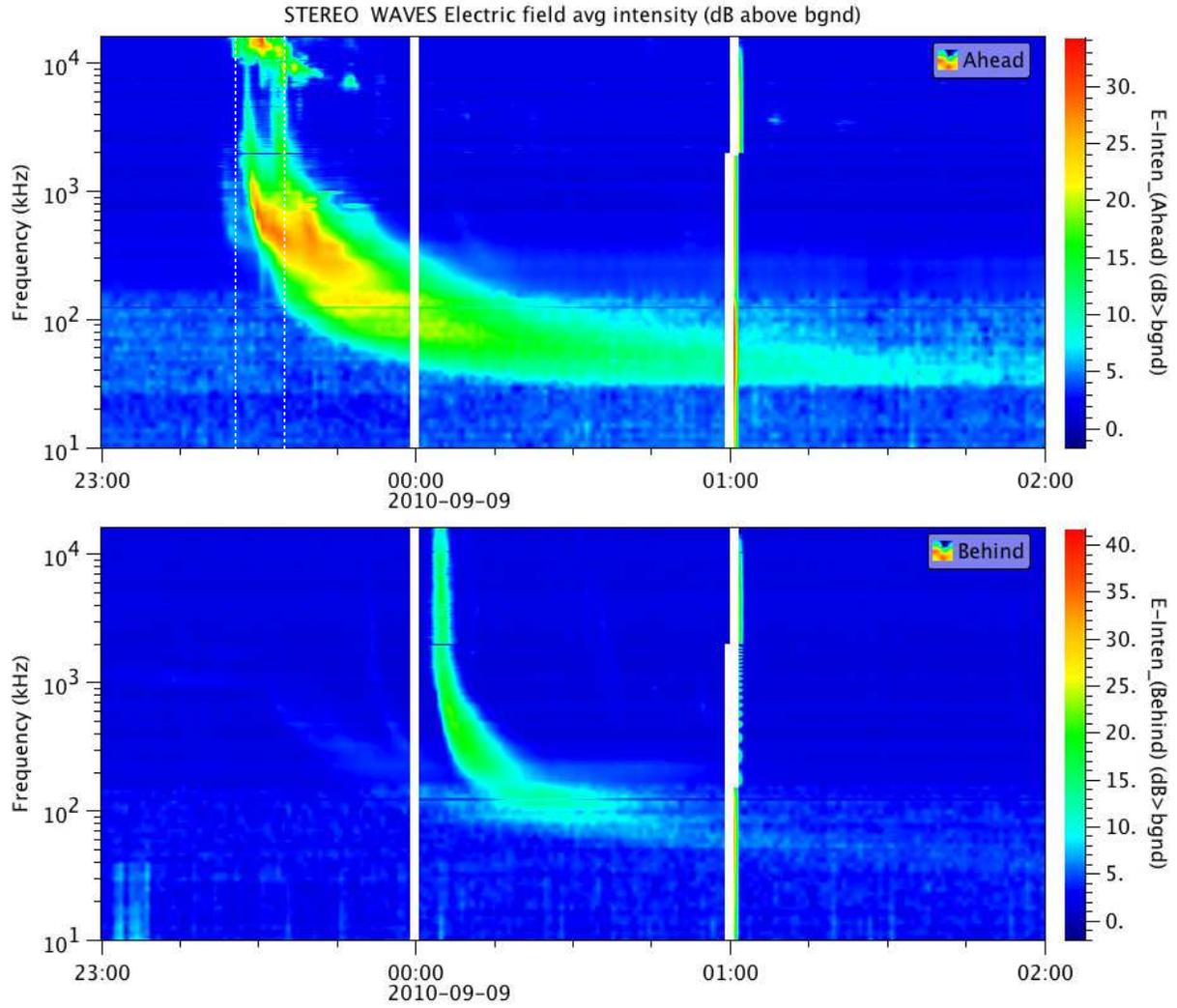}}
\caption{ Dynamic spectra of solar radio emission as measured by the WAVES instruments aboard (Upper) STEREO-A and (Lower) STEREO-B. The two vertical dashed lines in the STEREO-A panel at 23:25 and 23:35 UT, correspond to the effects from the first and second triggers, respectively, of the prominence oscillation as shown in Figure \ref{displ-curve}.}
\label{radio}
\end{figure}

The evidence for fast EUV wave could also be seen in STEREO WAVES observations. Figure~\ref{radio} shows radio  spectra from both STEREO/WAVES instruments. A strong radio emission is seen by STEREO-A, starting at 23:25 UT (single tone radio Type II burst, reported at frequencies from 16 to 8 MHz until 23:50 UT, see Interplanetary Type II/IV burst lists\footnote{http://ssed.gsfc.nasa.gov/waves/}). STEREO-B does not see corresponding emission as the active region lies on the far side in its field-of-view. The timings, corresponding to the two vertical dashed lines in Figure~\ref{displ-curve}(c),
are marked with two vertical dashed lines in the top panel of Figure~\ref{radio}. The fast EUV wave would lead to formation of MHD shocks in the  outer corona (16 MHz is the upper limit of the dynamic range of the SWAVES instrument and correspond to electron densities of $3 \times 10^6\, \rm{cm}^{-3}$, which are found at heights above $2 \textrm{R}_\odot$) leading to interplanetary Type II radio emission. The first vertical dashed line marks the onset of radio emission and also corresponds to the time when the displacement of the oscillating prominence reaches maximum amplitude.



We deduce that the arrival of the fast EUV wave at around 23:25 UT (first red arrow in Figure~\ref{euv-wave}) triggered the maximum displacement of the prominence (first vertical dashed line in Figure~\ref{displ-curve}(c)). While, the arrival of the slow ``EIT-like" wave at around 23:35 UT (second red arrow in Figure~\ref{euv-wave}) coincides well with the phase change of the prominence oscillation (second vertical dashed line) in Figure~\ref{displ-curve}(c). We propose that the two instants correspond to the interaction of the filament sheet with the fast and slow components of the EUV wave, respectively. While the fast EUV wave triggers the prominence oscillations the second one, i.e., the slow EUV wave arrives later and changes the phase of oscillation.
 The second (slow EUV) wave interacts with an already oscillating prominence. It can be understood as a second \lq{trigger\rq}, because it can reset or change the phase of the ongoing oscillation. This happens if the slow EUV wave is not in phase with the ongoing oscillation. Here the changing phase of the ongoing oscillation results in an apparent increase in the oscillation period, during the second oscillation cycle.


Finally, over a separation between fast and slow EUV wavefronts of 150 Mm, the slow EUV wave, travelling at a lower speed of $\sim250$ \kms, is expected to interact with the prominence with a delay of about 10 minutes. This delay is of the same order as the time difference between the effects of the primary and secondary triggers of the prominence oscillations, as observed in the oscillation curve (Figure \ref{displ-curve}). This strengthens the argument that the fast and slow EUV waves act as the primary and secondary triggers, respectively, for the prominence oscillation.

\subsection{Estimating the prominence magnetic field}

The coherent lateral displacement of the prominence structure suggests the collective oscillation mode to be a global kink mode (fundamental harmonic).
An MHD mode implies a characteristic frequency, which depends on the properties of the prominence rather than the triggering mechanism, as previously observed for large amplitude prominence oscillations \citep[e.g.][]{Ramsey1966,Hershaw2011}.
Applying the model of \cite{Kleczek1969} for  damped horizontal oscillations, the restoring force is due to magnetic tension
$$f_r=-\frac{1}{4\pi}(\textbf{B}\cdot\nabla)\textbf{B}~,$$
for an effective magnetic field \textbf{B} located in the axial direction of the prominence. Note that this theory assumes only one trigger. The characteristic period $P$ of this MHD mode of oscillation is independent of the trigger (global EUV wave types) and is given as
$$P=4\pi L B^{-1}\sqrt{\pi\rho_p}~,$$
 for an oscillating filament of length 2$L$, having mass density
$\rho_p$.

Taking $2L=400$ Mm $= 4.0 \times 10^{10}$ cm
and $\rho_p=10^{-14} g/cm^3$ and oscillation period of about 30
minutes we get a field strength estimate of $B=25$ Gauss,  which is of the order of the values inferred from coronal seismology \citep[see review by][ and references therein]{Nakariakov2005} as well as from spectro-polarimetric measurements \citep{Harvey2012}.
The application of prominence seismology requires accurate measurements of the characteristic period $P$. In the event studied, it is shown that the most reliable measurement of the characteristic period is at the start of the oscillation ($P_0$), which may be obtained from the data using a fitting function with variable period, as shown (or variable phase, not shown).

\section{Summary and Discussion}

We have analyzed the oscillations of a prominence sheet triggered by a nearby flare and observed with high spatial and temporal resolution from SDO/AIA. The unobscured view of the prominence oscillation could be obtained due to its favorable location with respect to the flaring site and to the Earth viewpoint. We summarize the results of the present analysis in the following points:

 \begin{itemize}
\item{ These high cadence observations allow us to derive the time-dependent period of the oscillations.  We use different methods to study the oscillation properties of the filament sheet, namely, the time-slice method, the local correlation tracking (LCT) method and the image cross-correlation method. All three methods give consistent values about the prominence oscillation period and amplitude.}

 \item{ We show that it is partly possible to apply LCT method to derive local velocity vectors of the oscillating structure (waveguide). The LCT velocity maps derived for the region outside the solar limb shows that the prominence structure oscillates coherently as a whole initially, suggesting a global MHD mode of oscillation. The amplitude of oscillation velocity near the top of the prominence reaches $\sim$11 \kms and is higher than that at the lower part of the prominence. Further, the LCT velocity vectors are mostly lateral  with respect to the prominence sheet, but do show a small inclination towards the solar limb.  This
may be the result of an apparent inclination in longitude due to the non-negligible solar heliographic latitude.
}

\item{  On the basis of the displacement profile of the prominence sheet derived from SDO/AIA images, space-time analysis of EUV wave using STEREO/EUVI images and the radio observations from STEREO/WAVES, it is observed that the prominence oscillations apparently undergo a phase change after the initial trigger by the fast EUV wave, which we interpret to be due to interaction with the slow EUV wave component. We propose a dual trigger of the prominence sheet as a result of its interaction with the fast and slow EUV wave components.
}

 \item{Further, assuming the simple model of \cite{Kleczek1969} we estimated the magnetic field  strength in the prominence to be of the order of $\approx$ 25 Gauss and highlighted the need for an accurate measurement of the characteristic period.}

\end{itemize}

 The present observations and analysis demonstrate that a secondary trigger by the slow EUV wave can influence the derived period and damping time of the oscillation. This phenomenon, which could be well-resolved and detected in a prominence due to the long-period, has more general implications for our understanding of damping mechanisms in the corona. The main implication we discuss is that the same dual trigger effect is expected to apply in coronal loops, although the comparison must be wary of the differences. Modes in coronal loops also exhibit collective transverse oscillations of the entire structure, albeit with different plasma parameters. Damping mechanisms for kink oscillations have been extensively studied in loops, and one can expect they are relevant to prominence, and vice-versa.
So far, there is no consensus on the exact damping mechanism at work, although there are two competing mechanisms that are closely scrutinized viz. resonant absorption \citep{Ionson1978,Hollweg1988,Ofman2002} and phase mixing \citep{Heyvaerts1983}.  According to \cite{Ofman2002}, one can distinguish between the damping mechanisms by comparing observations of periods ($P$) and damping times ($\tau$) with the scaling laws predicted by the two damping mechanisms. However, the observations show scatter and the two scaling laws are quite similar, i.e. $\tau\propto P$ for resonant absorption and $\tau \propto P^{4/3}$ for phase mixing, which makes the distinction difficult.

Using EIT images from SOHO, \cite{Hershaw2011} observed the large transverse oscillations of a prominence triggered twice by two different flare-associated coronal disturbances. Combining these results with those of previously analysed kink mode coronal loop oscillations, \citet{Hershaw2011}  extended the range of observed  periods and damping
times for kink oscillations (covering two orders of magnitude  in the parameter space of periods versus damping times), and found support for resonant absorption as the damping mechanism. The wide scattering of data points  (dispersion about the scaling law $\tau\propto P$) was considered
a result of the variation in density contrast and layer thickness over the numerous events \citep{Arregui2008,Hershaw2011}. Since the density contrast is much larger in prominence
than coronal loops, it was further suggested that the effect of this variation could be further reduced by the analysis of additional large amplitude transverse prominence oscillations \citep{Hershaw2011}.  \cite{Morton2009} studied the effect of radiative cooling on coronal loop oscillations and found that it can act as an efficient damping mechanism and can cause the period to decrease as a result. In the present case, we  have demonstrated that the oscillation period  may be seen to increase slightly due to the interaction with the slow EUV wave  and that the characteristic period $P_0$ could have been overestimated by $2.5\%$. This could affect our interpretations and efforts in constraining  the prominence magnetic field and the damping mechanisms at work.  We conclude that care should be taken when analyzing the observations of flare-induced oscillations in coronal structures such as prominence and coronal loops. In concrete terms, it would be important to reduce the scatter of data points and reproduce more accurate scaling laws by accounting for the dual trigger effect (i.e. phase change) when extracting periods and damping times in the analysis of displacement time series.

\acknowledgments
 The AIA data used here are courtesy of SDO (NASA) and the AIA consortium. The SECCHI data are courtesy of
STEREO and the SECCHI consortium. SOLIS data used here are produced cooperatively by NSF/NSO and NASA/LWS. National Solar Observatory (NSO) is operated by Association of Universities for Research in Astronomy (AURA, Inc) under a cooperative agreement with the NSF. C.F. acknowledges financial support from the UK Science and Technology Facilities Council (STFC) under her Advanced Fellowship.  Authors thank the anonymous referee for his/her useful suggestions and comments.

\bibliographystyle{apj}
\bibliography{agumnemonic,promg}

\end{document}